\documentstyle[sprocl]{article}
\input{psfig}
\def\Journal#1#2#3#4{{#1} {\bf #2}, #3 (#4)}
\def\prl{\em Phys. Rev. Lett.}
\def\prd{{\em Phys. Rev.} D}

\def\cqg{\em Class. Quantum Grav.}

\def\be{\begin{equation}}
\def\ee{\end{equation}}
\def\bea{\begin{eqnarray}}
\def\eea{\end{eqnarray}}

\begin{document}
\title{Dynamics of spin-2 fields in Kerr background geometries}
\author{ W. KRIVAN, P. LAGUNA, P. PAPADOPOULOS }
\address{
Dept. of Astronomy \& Astrophysics and\\
Center for Gravitational Physics \& Geometry\\
Penn State University, University Park, PA 16802 }
\maketitle\abstracts{We have developed a numerical method for evolving
perturbations of rotating black holes.
Solutions are obtained by integrating the Teukolsky equation
written as a first-order in time, coupled
system of equations, in a form that explicitly exhibits the radial
characteristic directions.
We follow the propagation of generic
initial data through the burst, quasi-normal ringing and power-law
tail phases. Future results may help to clarify the role of
black hole angular momentum on signals produced during the
final stages of black hole coalescence.
}
At first instance, a direct derivation of the equations governing
the perturbations of Kerr spacetimes is to consider perturbations of
the metric. This path, however, leads to gauge-dependent formulations.
A theoretically attractive alternative is to
examine {\em curvature} perturbations.  Using the
Newman-Penrose formalism, Teukolsky \cite{teuk72}
derived a master equation
governing not only gravitational perturbations (spin weight $s = \pm 2$) but
scalar, two-component neutrino and electromagnetic fields as well.
For the case $s=0$, it yields the equation for a scalar wave propagating in
a Kerr background, a system which we had studied previously \cite{PaperI}.

To our knowledge, most of the work on the dynamics of perturbations of
Kerr spacetimes has been performed under the assumption of a harmonic
time dependence.
Here we are interested in the time integration of physical initial
data, possibly from the inspiral collision of binary black holes. 
Fourier transformation of the data and subsequent evolution of such
data in the frequency domain approach is, in principle, 
possible but numerically cumbersome. The main complication one faces by keeping
the equation in the time domain is that one cannot longer benefit from the
reduction of dimensionality implied by the separation of variables.
We have chosen the option to evolve one single 2+1
PDE instead of the equivalent approach of solving the set of ODEs 
corresponding to the Fourier spectrum. The computational burden of both
approaches is likely similar.  
The resulting evolution equation is a hyperbolic, linear equation
which is quite amenable to numerical treatment, provided suitable
coordinates, variables and boundary conditions are chosen.

The two key factors in successfully solving the Teukolsky equations were:
first, to carefully
select the evolution field and its asymptotic behavior, and second 
to rewrite the Teukolsky equation in a form that explicitly
exhibits the radial characteristic directions.
On the analytical level, one obtains bounded solutions for any 
direction of propagation by choosing $s=-2$ and rescaling by an 
appropriate function of
$r$. A convenient choice is simply $r^3$, a factor that is regular at
the horizon. 
Regarding the choice of spatial coordinates,
we use the Kerr tortoise
coordinate $r^*$ and the Kerr $\tilde\phi$ coordinate 
instead of the Boyer-Lindquist coordinate $\phi$.
Then the ansatz for the solution to the
Teukolsky equation is:
$
\Psi(t,r^*,\theta,\tilde\phi) \equiv e^{i m
\tilde\phi}\,r^3\,\Phi(t,r^*,\theta)$ .
After a series of unsuccessful numerical experiments with this
second-order in time and space equation for $\Phi$,
we found that numerical instabilities due to the first order in time
derivatives in the Teukolsky equation were suppressed by
introducing an auxiliary field $\Pi \equiv \partial_t{\Phi} + b \,
\partial_{r^*}\Phi $, that converts the Teukolsky
equation to a coupled set of first-order equations in space and time
\cite{PaperII}, where
$b  \equiv ( {r}^{2}+{a}^{2})/ \Sigma$ and
$\Sigma^2 \equiv   (r^2+a^2)^2-a^2\,\Delta\,\sin^2\theta$.
The resulting first order system is hyperbolic
in the radial direction.
Stable evolutions were achieved using
a modified Lax-Wendroff method,
discretizing the equation on a uniform two dimensional polar
grid. 
Typically we used a computational domain of size $-100M \le r_i^* \le 500M$
and $0 \le \theta_j \le \pi$ with $0 \leq i \leq 8000$ and 
$0 \leq j \leq 32$. Details of the numerical
scheme are described in \cite{PaperII}.
The stability of the code was verified with long-time evolutions,
of the order of $1000M$. Its accuracy in turn was tested using
standard convergence tests.

\begin{figure} 
\hspace*{2.5cm}
\psfig{figure=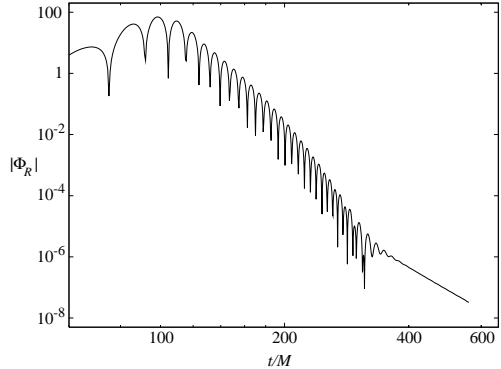,height=2.0in} 
\caption[figure1]{\label{figure1} 
Log-log plot of the time evolution of the field amplitude
$|\Phi_R|$ for an observer located at $r^*=20M$, $\theta = \pi/2$ 
($a=0.9M$, $m=0$). 
The exponentially damped oscillation of \\
\hspace*{-2.3cm} the
scattered signal, and the late time power-law tail are clearly
exhibited.
}
\end{figure}

\vspace{0.1cm}
The evolution of generic initial data consists of three stages, as seen
from an observer located away from the hole.  During the first stage,
the observed signal depends on the structure of the initial pulse and
its reflection from the angular momentum barrier (burst phase). This
phase is followed by an exponentially decaying quasinormal ringing of
the black hole (quasinormal phase). In the last stage, the wave
slowly dies off as a power-law tail (tail phase).
The numerical algorithm described in the previous section is used
to obtain the time evolution of generic perturbations impinging
on the rotating black hole. The broad features of the evolution
are demonstrated in Fig.~\ref{figure1}, where quasinormal ringing and tail
behavior are clearly manifested.
%
%
We then performed a series of simulations with $m=0$ for different values of
$a$. Focusing on the later part of the ringing
sequence, which clearly depicts the least damped mode of
oscillation, we read off the oscillation frequencies.
A comparison of our values with those
given by Kokkotas \cite{kostas} and Leaver \cite{leaver}
shows an agreement of better than $1\%$.

We have shown \cite{PaperI}, that the late time evolution of scalar
fields in the background of rotating black holes is qualitatively
similar to the non-rotating case. We extend here this result to
the physically more interesting spin-2 field evolution.
Our calculations show that the exponents governing the behavior for
$a \ne 0$ do not exhibit a significant change when
compared to the Schwarzschild case, if the initial data pulse is 
not given by the lowest allowed mode for a particular value of $m$.
That is, the power-law tail behavior is basically determined by 
the dominant asymptotic form of the potential.
When the initial data pulse is not given by the
lowest allowed mode for a particular value of $m$, mixing
of modes occurs \cite{PaperI}.

\vspace{0.1cm}
We thank  H.-P.\ Nollert, R.\ Price and J.\ Pullin for
helpful discussions. This work was supported by the Binary Black Hole
Grand Challenge Alliance, NSF PHY/ASC 9318152 (ARPA supplemented) and
by NSF grants PHY 96-01413, 93-57219 (NYI) to PL. WK was
supported by the Deutscher Akademischer Austauschdienst (DAAD).

\end{document}